\documentclass{aa}  

\usepackage{graphicx}
\usepackage{txfonts}
\usepackage{color}
\usepackage{stfloats}
\usepackage{multirow}
\usepackage{mhchem}
\usepackage{booktabs}
\usepackage{hyperref}

\newcommand{\jms}{J. Mol. Spectr.}
\newcommand{\jmst}{J. Mol. Struct.}

\newcommand{\natastro}{Nat. Astron.}
\newcommand{\pccp}{Phys. Chem. Chem. Phys.}

\newcommand{\science}{Science}

\begin{document}

\title{Low D/H ratio for benzonitrile in \mbox{TMC-1}: Implication for the origin of polycyclic aromatic hydrocarbons in cold dark clouds\thanks{Based on observations carried out with the Yebes 40m telescope (projects 19A003, 20A014, 20D023, 21A011, 21D005, and 23A024). The 40m radio telescope at Yebes Observatory is operated by the Spanish Geographic Institute (IGN; Ministerio de Transportes, Movilidad y Agenda Urbana).}}

\titlerunning{Low D/H ratio for benzonitrile in \mbox{TMC-1}}
\authorrunning{Steber et al.}

\author{
A.~L.~Steber\inst{1},
J.~Janeiro\inst{2},
C.~Cabezas\inst{3},
M.~Ag\'undez\inst{3},
M.~Pereira-Santaella\inst{3},
C.~P\'erez\inst{1},
D.~P\'erez\inst{2},
D.~Heras\inst{1},
A.~Lesarri\inst{1},
I.~Garc\'ia-Bernete\inst{4},
J.~R.~Goicoechea\inst{3},
and
J.~Cernicharo\inst{3}
}


\institute{
Departamento de Qu\'imica Física y Qu\'imica Inorg\'anica, Facultad de Ciencias-I.U. CINQUIMA, Universidad de Valladolid, E-47011 Valladolid, Spain\\
\email{amanda.steber@uva.es}
\and
Centro Singular de Investigaci\'on en Qu\'imica Biol\'oxica e Materiais Moleculares (CiQUS) and Departamento de Qu\'imica Org\'anica, Universidade de Santiago de Compostela, E-15782 Santiago de Compostela, Spain\\
\email{dolores.perez@usc.es}
\and
Departamento de Astrof\'{i}sica Molecular, Instituto de F\'{i}sica Fundamental, CSIC, Calle Serrano 123, E-28006 Madrid, Spain\\\email{marcelino.agundez@csic.es}
\and
Centro de Astrobiolog\'ia (CAB), CSIC-INTA, Camino Bajo del Castillo s/n, E-28692 Villanueva de la Ca\~nada, Madrid, Spain
}

\date{Received; accepted}

 
\abstract
{Radioastronomical observations have recently discovered polycyclic aromatic hydrocarbons (PAHs) of moderate size (up to 24 carbon atoms) in cold dark clouds, although it is currently unknown whether they are formed in situ through a bottom-up mechanism or from larger PAHs (20-100 carbon atoms) inherited from a previous diffuse stage in a top-down scenario. Infrared observations have recently shown that large PAHs present in UV-illuminated regions are strongly enriched in deuterium. In order to shed light on the origin of PAHs in cold clouds, we have searched for deuterated benzonitrile in the cold dark cloud \mbox{TMC-1}. To that purpose we have synthesized the three isomers (\textit{ortho}, \textit{meta}, and \textit{para}) of monodeuterated benzonitrile, measured their rotational spectra across the 2-18 GHz and 75-110 GHz frequency ranges in the laboratory, and searched for them in \mbox{TMC-1} using data from the QUIJOTE line survey. We did not detect any of the three species and have derived a 3$\sigma$ upper limit on the column density of each of them of 3.0\,$\times$\,10$^{10}$ cm$^{-2}$, meaning a fractional abundance relative to H$_2$ of $<$\,3\,$\times$\,10$^{-12}$. We derived a D/H ratio (which we define as the total number of D atoms with respect to the total number of H atoms present in benzonitrile) of $<$\,1.2\,\%. This value is in line with the range of D/H ratios observed for other molecules in \mbox{TMC-1} (0.06-3.3\,\%), where deuterium enrichment is explained in terms of isotopic fractionation at low temperature. It is however below the range of D/H ratios derived for large unspecific PAHs from JWST observations of the galactic photodissociation regions (PDRs) Orion Bar and M\,17 and the galaxies M\,51 and NGC\,3256-S (between 1\,\% and $<$\,17\,\%). Although it is not straightforward to compare the deuteration of PAHs in dark and UV-irradiated clouds, our results suggest that the population of PAHs detected in cold dark clouds does not result from the fragmentation of larger PAHs inherited from the previous diffuse stage in a top-down scenario.}


\maketitle

\section{Introduction}

One of the most surprising discoveries made in recent years in astrochemistry has been the discovery of abundant polycyclic aromatic hydrocarbons (PAHs) in cold dark clouds. Molecules with one ring or several fused rings have been unambiguously identified in the well-known dark cloud Taurus Molecular Cloud 1 (\mbox{TMC-1}). These comprise derivatives of cyclopentadiene, benzene, indene, naphthalene, acenaphthylene, pyrene, and coronene \citep{Cernicharo2021a,Cernicharo2021b,Cernicharo2024a,McGuire2018,McGuire2021,McCarthy2021,Lee2021,Burkhardt2021a,Wenzel2024,Wenzel2025}. Moreover, some of these cycles have also been detected in other dark clouds similar to \mbox{TMC-1} \citep{Burkhardt2021b,Agundez2023a}.

It is nowadays a matter of debate whether these PAHs are synthesized in situ in the cold dark clouds where they are observed through a bottom-up mechanism \citep{Cernicharo2021b,Byrne2024} or whether they result from the fragmentation of larger PAHs inherited from a previous evolutionary stage in a top-down scenario \citep{Pety2005,Zhen2014,Burkhardt2021b,Goicoechea2025}. Indeed, large PAHs with estimated sizes of 20-100 carbon atoms are inferred to be present in evolutionary stages earlier than that of cold dark clouds, such as in diffuse clouds, through the observation of so-called aromatic infrared bands \citep{Tielens2008,Sandstrom2023}. The observation of emission bands at 4.65 $\mu$m and 4.35 $\mu$m in some photodissociation regions (PDRs) bright in PAH emission indicates that interstellar PAHs are strongly enriched in deuterium \citep{Peeters2004,Peeters2024,Onaka2014,Doney2016,Boersma2023,Pereira-Santaella2024,Yang2025,Draine2025}. This fact offers a way to test the top-down scenario for the origin of PAHs in cold dark clouds. If the PAHs identified in dark environments arise from larger PAHs inherited from a previous UV-illuminated phase, we would expect them to show this enrichment in deuterium.

It is, however, difficult to put constraints on the abundance of deuterated PAHs in cold clouds because the cyano derivatives of indene, naphthalene, acenaphthylene, pyrene, and coronene are observed through relatively weak lines in \mbox{TMC-1}, and thus the lines from the deuterated forms are most likely below the noise level of currently available spectra. Benzonitrile (C$_6$H$_5$CN), however, is observed through relatively intense lines in \mbox{TMC-1} \citep{Cernicharo2021b} and, although it is not formally a PAH, it can be seen as a member of the family of aromatic rings. The search for deuterated benzonitrile in cold dark clouds is therefore feasible. We have characterized the rotational spectra of the three mono-deuterated isomers of benzonitrile, for which deuterium is substituted at the \textit{ortho}, \textit{meta}, and \textit{para} positions, in the laboratory and searched for them in \mbox{TMC-1} using the QUIJOTE (Q-band Ultrasensitive Inspection Journey to the Obscure TMC-1 Environment) line survey carried out with the Yebes\,40m telescope \citep{Cernicharo2021c}. In Sect.\,\ref{sec:spectroscopy} we present the laboratory measurements of the rotational spectra of \textit{ortho}, \textit{meta}, and \textit{para} monodeuterated benzonitrile, in Sect.\,\ref{sec:observations} we describe the astronomical search for deuterated benzonitrile in \mbox{TMC-1}, in Sect.\,\ref{sec:discussion} we discuss the implications for the origin of PAHs in cold dark clouds, and in Sect.\,\ref{sec:conclusions} we summarize our conclusions.

\section{Laboratory rotational spectrum of deuterated benzonitrile} \label{sec:spectroscopy}

\subsection{Microwave measurements} \label{sec:experimental_mw}

\begin{figure*}
\centering
\includegraphics[angle=0,width=0.85\textwidth]{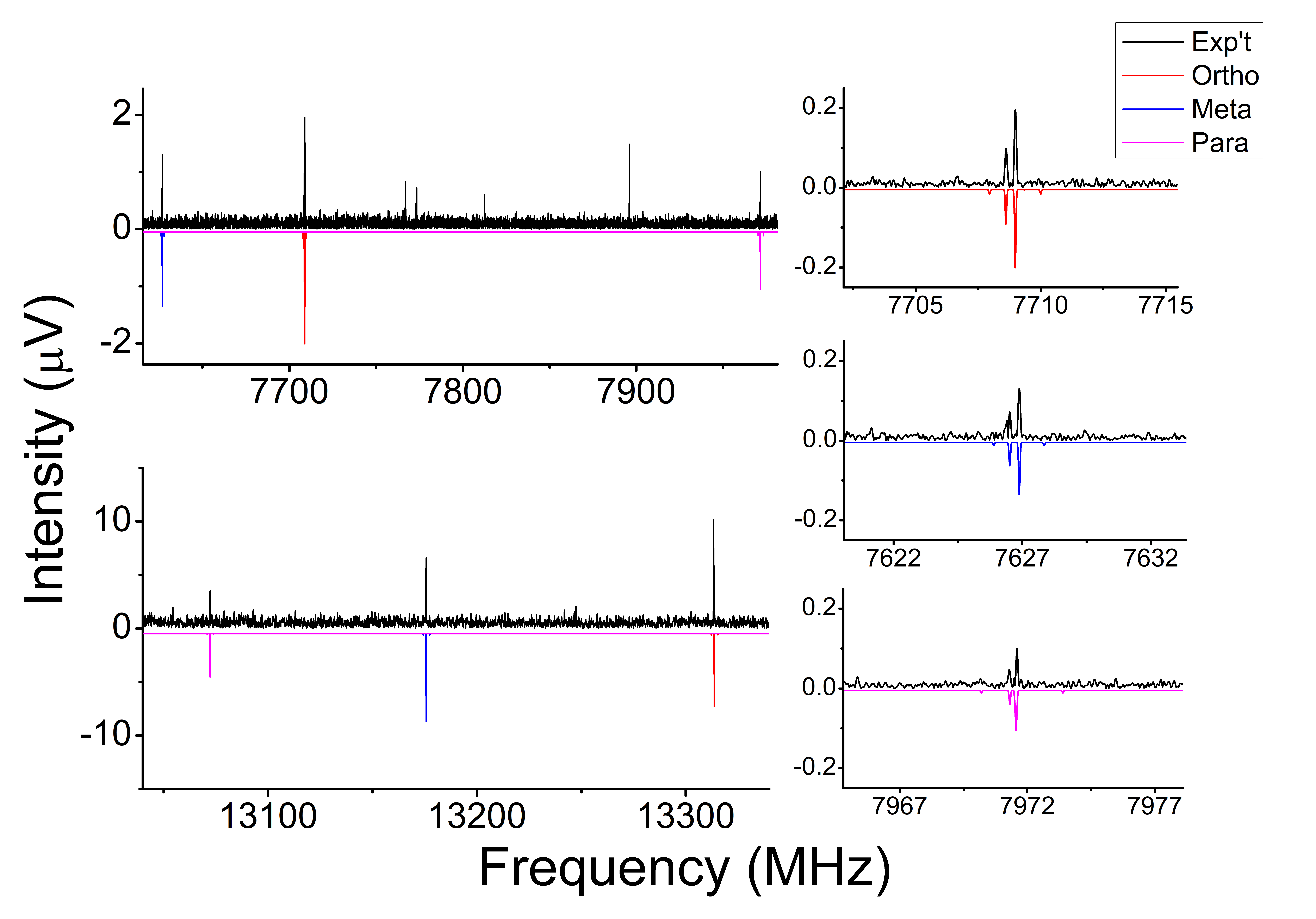}
\caption{Insets of the recorded spectrum in the discharge of deuterated benzene and acetonitrile. The three species of deuterated benzonitrile (\textit{ortho}, \textit{meta}, and \textit{para}) can be seen. The experimental spectra are shown in black above the x axis, while the simulated spectra for each isomer at 1 K (derived from the rotational parameters) are shown in colors below the x axis.} \label{fig:mw}
\end{figure*}

The rotational spectra of the three isomers of deuterated benzonitrile were measured in several steps across the 2-18 GHz frequency range using an electrical discharge source \citep{Ohshima1992,McCarthy2000}, which allowed us to generate in situ the three isomers of deuterated benzonitrile. A sample of $\geq$ 97\% deuterated benzene was purchased from Sigma Aldrich and used without further purification. This was used to make a 0.5\% gas sample diluted in neon, and it was then passed over an external reservoir containing acetonitrile (CH$_3$CN). The mixture was passed through a modified Parker series 9 valve, followed by the electrical discharge nozzle, and a subsequent supersonic expansion into the vacuum chamber with a pressure of 3 bar. The experimental conditions for the discharge were 1 kV and 100 mA. This resulted in the production of the \textit{ortho}, \textit{meta}, and \textit{para} isomers of deuterated benzonitrile. For the measurements in the 2-8 GHz range, the microwave setup has been described previously \citep{Moran2025}, and the operation allows the entire range to be acquired in each acquisition. A chirped-pulse is generated by an arbitrary waveform generator, then amplified and broadcast across the chamber with a broadband horn antenna, operating in the 2-18 GHz range. The chirped-pulse interacts with the molecular signal, after which a free induction decay is produced. The free induction decay is collected with a similar horn antenna as the broadcasting antenna, amplified, and digitized on a Tektronix DPO 70804C oscilloscope. In the 8-18 GHz region, due to power limitations, targeted measurements based on the predicted transitions of the three isomers were performed. The setup for this frequency range operated in much the same way as that in the 2-8 GHz range, with some key differences. The oscilloscope was changed to a Tektronix DPO 72004, which allowed us to use a sampling rate of 50 GS/s with a maximum bandwidth of 20 GHz. The chirped-pulse amplification was carried out by a solid state amplifier (Microsemi AML218P4013) and the low-noise amplifier was substituted with a 2-18 GHz low-noise amplifier (Miteq LNAS-55-01001800-22-10P).  Each measurement was 1 GHz in length, 100 000 acquisitions were acquired, and a 10 $\mu$s free induction decay was collected. In both setups, the fast frame feature of the Tektronix oscilloscopes was used to increase the overall effective repetition rate. Insets of the generated spectra can be seen in Fig.\,\ref{fig:mw}.

\subsection{Millimeter-wave measurements} \label{sec:experimental_mmw}

\begin{figure*}
\sidecaption
\includegraphics[angle=0,width=12cm]{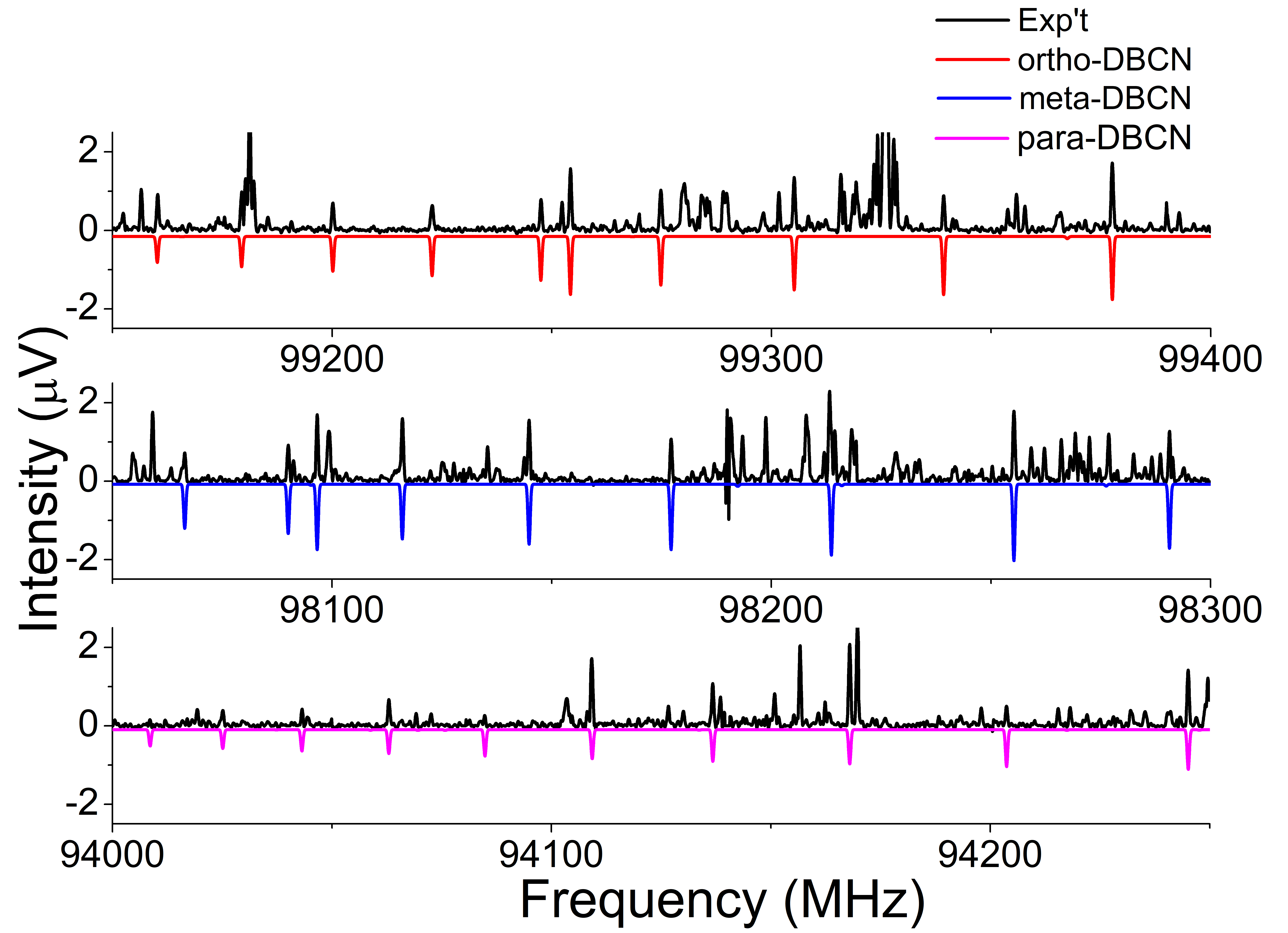}
\caption{Portion of the spectra of \textit{ortho}- (top), \textit{meta}- (middle), and \textit{para}- (bottom) deuterated benzonitrile (DBCN) recorded in the 75-110 GHz range. The experimental spectra are shown in black above the x axis, while the simulated spectra at 300 K (calculated from the rotational parameters) are shown in colors below the x axis.} \label{fig:mmw}
\end{figure*}

Measurements of the rotational spectra of the three isomers of deuterated benzonitrile were also carried out at millimeter wavelengths (75-110 GHz), using in this case samples synthesized in the laboratory. The synthesis was achieved by palladium-catalyzed bromo-deuterium exchange \citep{Zhang2015}, starting from the commercially available \textit{ortho}-, \textit{meta}- and \textit{para}-bromobenzonitrile, respectively, and using sodium formate-D (DCOONa) as a deuterium anion source. The three deuterium-labelled benzonitriles were isolated in reasonable yields (56-70\,\%), as is shown by gas chromatography mass spectrometry (GC-MS). The synthesis of deuterated benzonitrile in described in detail in Appendix \ref{app:synthesis}.

We used a W-band instrument purchased from BrightSpec. Inc. to record the millimeter wave spectra. This instrument is equipped with a 66 cm long stainless steel cell that has a diameter of 6.5 cm and is closed on either end with teflon lenses \citep{Zaleski2017,Arenas2017}. It can be used in a static or flow cell configuration. Also, this instrument is able to operate in either fast mode, in which larger segments can be measured and stitched together (decreasing the measurement time), or in high dynamic range mode, in which smaller bandwidth segments are used to acquire the entire frequency range (reducing the spurious signal content due to the upconversion of the original chirped-pulse). Both modes are an implementation of segmented chirped-pulse spectroscopy \citep{Neill2013}. In this experiment, we used the static cell configuration and the high dynamic range mode to acquire the three spectra. Each of the synthesized samples was transferred from the original vial into a Schlenk flask. However, due to the low quantity, a few drops of commercially available benzonitrile were added to the samples to facilitate this process. The chamber was closed off from vacuum and approximately 2.5 mTorr of vapor was added to the cell. Due to outgassing from the walls of the chamber, the pressure rose during the course of the measurement to somewhere between 10-15 mTorr for each isomer. An inset of the recorded spectra at room temperature can be found in Fig.\,\ref{fig:mmw}.

\subsection{Analysis of the rotational spectra} \label{sec:analysis}

The assignment of the rotational spectra of the three species was straightforward as the rotational transition frequencies with low $K_{a}$ values were well predicted, within a few megahertz (in the millimeter wave region), by using the rotational constants reported by \cite{Bak1962} and \cite{Casado1971}. As was stated before, we measured rotational transitions for the three isomers in the microwave and millimeter wave regions. Some of the transitions in the microwave region show nuclear quadrupole coupling hyperfine splittings produced by $^{14}$N, which has a nuclear spin of $I$ = 1 (see Fig.\,\ref{fig:mw}). Even though deuterium also has a nuclear spin of $I$ = 1, its nuclear quadrupole coupling effects are much smaller than the ones induced by the $^{14}$N nucleus. For the millimeter wave region, all the transitions were observed as single lines since these nuclear quadrupole coupling hyperfine splittings are smaller than the experimental broadening of the lines (see Fig.\,\ref{fig:mmw}). A total of 657, 628, and 592 rotational transitions were observed for the \textit{ortho}, \textit{meta}, and \textit{para} species of deuterated benzonitrile, respectively. A list with all the measured frequencies is provided (see Data Availability section).

For each isomer, all the observed transitions were analyzed in a combined fit using the SPFIT program \citep{Pickett1991} with the $A$-reduction of the Watson's Hamiltonian in $I^r$ representation \citep{Watson1977}. The analysis rendered the experimental molecular constants listed in Table \ref{constants}. As can be seen, all the quartic centrifugal constants could be accurately determined, except for $\Delta_K$, which was kept fixed to the value determined for benzonitrile by \citet{Zdanovskaia2018}. In the same manner, we fixed the sextic centrifugal constants $\Phi_J$, $\Phi_{JK}$, $\Phi_{KJ}$, $\Phi_K$, $\phi_J$, and $\phi_K$ and octic centrifugal constants $L_J$, $L_{JJK}$, $L_{KKJ}$, and $L_K$ to the values determined for benzonitrile by \cite{Zdanovskaia2018} for all three isomers. Thanks to the observation of rotational transitions in the microwave range, we could derive the values of the nuclear quadrupole coupling constants $\chi_{aa}$ and $\chi_{bb}$, except in the case of the \textit{meta} species where only $\chi_{aa}$ could be determined. In this case, $\chi_{bb}$ was fixed to the value of benzonitrile \citep{Zdanovskaia2018}.

\begin{table}
\begin{center}
\small
\caption[]{Spectroscopic parameters of the three monodeuterated isomers of benzonitrile.}
{\label{constants}
\begin{tabular}{lccc}
\hline
\hline
Constants  & \textit{ortho} & \textit{meta} & \textit{para} \\
\hline
$A$ (MHz)               &  5379.4134(48)$^a$    &  5383.7464(52)    &   5655.02893(72)  \\
$B$ (MHz)               &   1546.11371(26)      & 1526.28037(27)    &   1496.59469(12)  \\
$C$ (MHz)               &   1200.71610(23)      & 1188.93224(24)    &   1183.20832(11)  \\
$\Delta_J$ (kHz)        &    0.044862(59)       &  0.043273(61)     &    0.041982(35)   \\
$\Delta_{JK}$ (kHz)     &     0.92203(26)       &   0.89208(28)     &    0.84428(11)    \\
$\Delta_K$ (kHz)        &    [0.24234]$^b$      &   [0.242343]      &     [0.242343]    \\
$\delta_J$ (kHz)        &    0.011279(43)       &  0.010756(45)     &    0.009957(14)   \\
$\delta_K$ (kHz)        &     0.5951(19)        &   0.5723(21)      &    0.55402(69)    \\
$\chi_{aa}$ (MHz)       &     $-$4.210(41)        &    $-$4.21(12)      &    $-$4.18(13)    \\
$\chi_{bb}$ (MHz)       &      2.169(62)        &    [2.28872]      &      2.30(46)   \\
\\
$N_{lines}$           &         657           &       628         &        592        \\
$\sigma_{rms}$ (kHz)    &        23.5           &      24.2         &        24.5       \\
$J$     &        1-45           &      1-45         &        2-44       \\
$K_a$ &        0-34           &      0-31         &        0-35       \\
\hline
\end{tabular}
}
\end{center}
\tablefoot{$^a$\,Numbers in parentheses represent the derived uncertainty (1\,$\sigma$) of the parameter in units of the last digits. $^b$ Values in brackets have been fixed to the ones derived by \citet{Zdanovskaia2018} for benzonitrile.}
\end{table}

Using the molecular constants from Table \ref{constants} we obtained accurate predictions of the rotational transition frequencies for the three isomers in the Q band, with uncertainties of less than 10 kHz for $a$-type $R$-branch transitions with $K_{a}$ values <10. This represents an increase by a factor $\sim$50 in the accuracy of the predictions in the Q band. For the intensity predictions, it was assumed that the three deuterated forms have the same dipole moment along the $a$ axis that was measured for the parent species, 4.5152\,$\pm$\,0.0068 D \citep{Wohlfart2008}. The rotational partition functions employed for each isomer are shown in Table \ref{pfunction}. They were calculated using the SPCAT program \citep{Pickett1991} at a maximum value of $J$ = 100 and not including the hyperfine constants.

\section{Astronomical observations: Search for deuterated benzonitrile in \mbox{TMC-1}} \label{sec:observations}

Astronomical data from the QUIJOTE survey carried out with the Yebes 40m telescope \citep{Cernicharo2021c} were used to search for monodeuterated benzonitrile in \mbox{TMC-1}. Briefly, QUIJOTE consists of a Q-band line survey (31.0-50.3 GHz) of the cold dark cloud \mbox{TMC-1} at the position where cyanopolyyne emission peaks ($\alpha_{J2000}=4^{\rm h} 41^{\rm  m} 41.9^{\rm s}$ and $\delta_{J2000}=+25^\circ 41' 27.0''$). The observations are carried out using the frequency-switching technique, with a frequency throw of either 8 or 10 MHz. The whole Q band is covered in one shot with a spectral resolution of 38.15 kHz in horizontal and vertical polarizations using a 7 mm dual linear polarization receiver connected to a set of 2\,$\times$\,8 fast Fourier transform spectrometers \citep{Tercero2021}. The intensity scale at the Yebes\,40m telescope is the antenna temperature, $T_A^*$, which has an estimated uncertainty due to calibration of 10\,\%. The antenna temperature can be converted to main beam brightness temperature, $T_{\rm mb}$, by dividing $T_A^*$ by $B_{\rm eff}$/$F_{\rm eff}$. The beam efficiency, $B_{\rm eff}$, is given by the Ruze formula $B_{\rm eff}$\,=\,0.797\,$\exp{[-(\nu/71.1)^2]}$, where $\nu$ is the frequency in GHz, and the forward efficiency, $F_{\rm eff}$, is 0.97, as measured at the Yebes\,40m telescope. The half power beam width (HPBW) can be approximated as HPBW($''$) = 1763/$\nu$, where $\nu$ is the frequency in gigahertz. Observations were carried out from November 2019 to July 2024 with a total on-source telescope time of 1509.2 h, of which 736.6 h correspond to a frequency throw of 8 MHz and 772.6 h to a throw of 10 MHz. The $T_A^*$ root mean square (rms) noise level varies between 0.06 mK at 32 GHz and 0.18 mK at 49.5 GHz. The data were analyzed using the GILDAS software\footnote{\texttt{https://www.iram.fr/IRAMFR/GILDAS/}} following the procedure described in \cite{Cernicharo2022}.

\begin{table}
\small
\begin{center}
\caption[]{Rotational partition function of the three monodeuterated isomers of benzonitrile at different temperatures.}
\label{pfunction}
\begin{tabular}{lrrr}
\hline
\hline
Temperature (K)  & \multicolumn{1}{c}{\textit{ortho}} & \multicolumn{1}{c}{\textit{meta}} & \multicolumn{1}{c}{\textit{para}} \\
\hline
300  &   242037.0   &  243898.5  &  240355.3 \\
225  &   169059.2   &  170581.1  &  168227.7 \\
150  &    96627.8   &   97627.4  &   96373.9 \\
 75  &    34658.5   &   35040.4  &   34609.4 \\
 37.5  &    12258.0   &   12393.3  &   12241.1 \\
 18.75  &     4335.8   &    4383.7  &    4329.8 \\
  9.375  &     1534.4   &    1551.4  &    1532.3 \\
\hline
\end{tabular}
\end{center}
\end{table}

\begin{figure*}
\centering
\includegraphics[angle=0,width=0.95\textwidth]{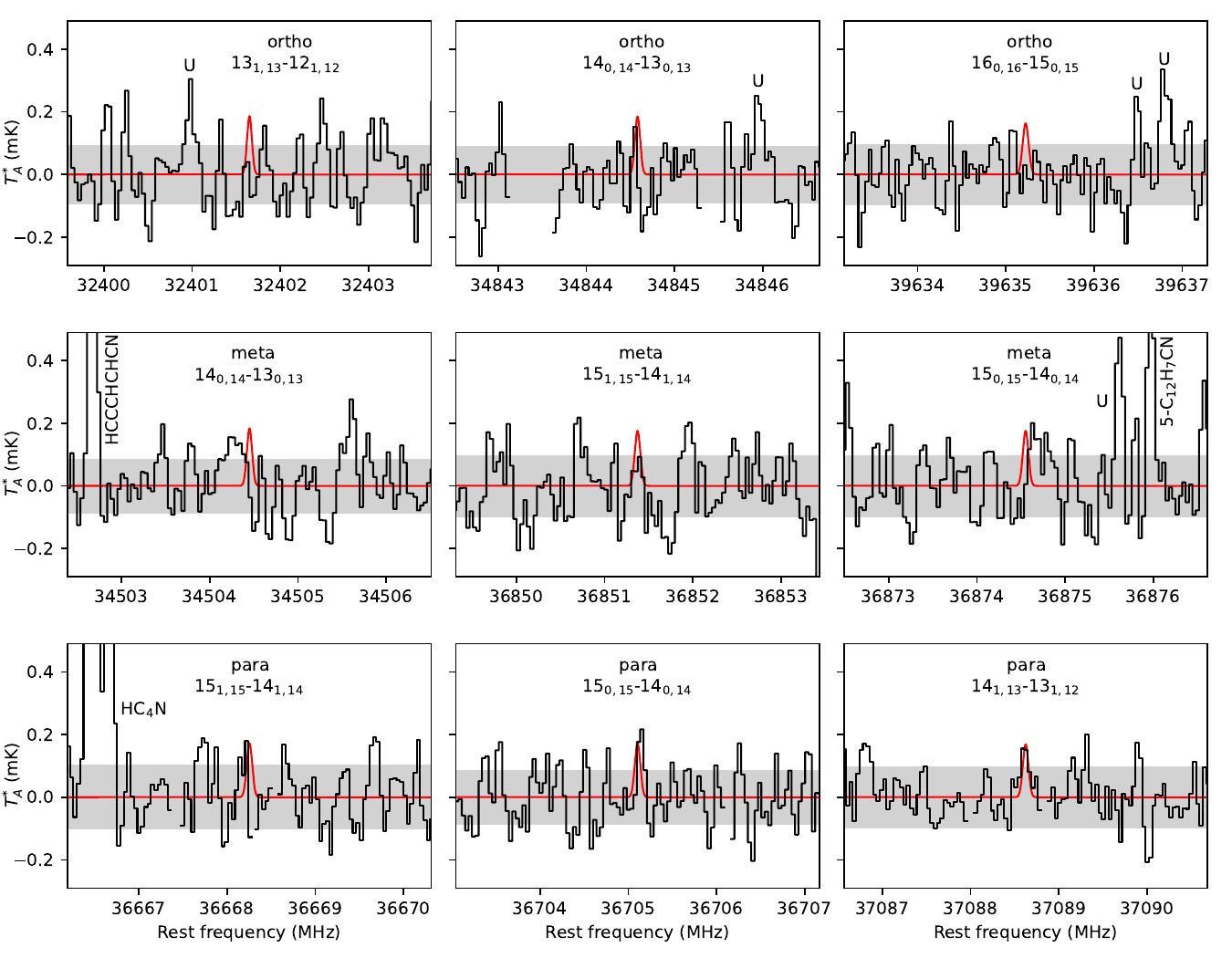}
\caption{Spectra of \mbox{TMC-1} in the Q band at the frequencies of some of the most favorable lines of \textit{ortho}, \textit{meta}, and \textit{para} deuterated benzonitrile. Negative artifacts produced by the frequency-switching technique have been blanked. The noise level, measured in a window of $\pm$\,8 MHz around the expected position of each line with the nominal spectral resolution of 38.15 kHz, is indicated by a horizontal gray band. The red lines correspond to the line intensities calculated adopting the 3$\sigma$ upper limits on the column densities derived here and a full width at half maximum of 0.60 km s$^{-1}$.} \label{fig:lines}
\end{figure*}

The three deuterated isotopologs of benzonitrile have only $a$-type rotational transitions. At a rotational temperature of 9.0 K, which is the gas kinetic temperature \citep{Agundez2023b} and the rotational temperature of the parent species of benzonitrile in \mbox{TMC-1} \citep{Cernicharo2021b}, the most favorable lines in the Q band correspond to rotational transitions with $K_a$\,=\,0 and 1. The $K_a$\,=\,0 and 1 lines of the three isomers of deuterated benzonitrile lying in the Q band have a modest hyperfine splitting ($<$\,20 kHz). Lines with $K_a$\,$\ge$\,2 have larger hyperfine splittings (30-50 kHz), similar to the values observed for the non-deuterated species in \mbox{TMC-1} \citep{Cernicharo2021b}. We searched for the $K_a$\,=\,0 and 1 lines of deuterated benzonitrile in our QUIJOTE data, but there is no clear evidence of them. In Fig.\,\ref{fig:lines} we show the spectra at the frequencies of three selected transitions for each of the three deuterated isotopologs, \textit{ortho}, \textit{meta}, and \textit{para}, of benzonitrile. In order to derive upper limits on the column densities of the three deuterated species, we adopted a rotational temperature of 9.0 K, derived from the analysis of 100 individual lines of the parent species of benzonitrile in \mbox{TMC-1} by \cite{Cernicharo2021b}, and assumed that the emission of these species would be distributed as a circle with a radius of 40\,$''$, which is approximately the size measured for the parent species of benzonitrile in \mbox{TMC-1} \citep{Cernicharo2023}. We derived a 3$\sigma$ upper limit on the column density of \textit{ortho}, \textit{meta}, and \textit{para} deuterated benzonitrile of 3.0\,$\times$\,10$^{10}$ cm$^{-2}$. Adopting a column density of H$_2$ of 10$^{22}$ cm$^{-2}$ \citep{Cernicharo1987}, the fractional abundance of each of the three deuterated isomers relative to H$_2$ is $<$\,3\,$\times$\,10$^{-12}$. The calculated line profiles adopting the 3$\sigma$ upper limit derived as the column density and a full width at half maximum of 0.60 km s$^{-1}$ \citep{Agundez2023b} are shown in Fig.\,\ref{fig:lines}. The column density derived for the parent species of benzonitrile in \mbox{TMC-1} is 1.2\,$\times$\,10$^{12}$ cm$^{-2}$ \citep{Cernicharo2021b}. Therefore, the abundance ratio of deuterated to non-deuterated benzonitrile is  $<$\,2.5\,\% for the \textit{ortho}, \textit{meta}, and \textit{para} forms.

\section{Discussion} \label{sec:discussion}

The D/H ratio of a given molecule is defined here as the total number of D atoms contained in the different deuterated versions of the molecule divided by the total number of H atoms present in the non-deuterated and deuterated forms of that molecule. This definition applies to small individual molecules observed at radio frequencies, such as benzonitrile, and to large unspecific PAHs observed in the infrared, in which case the D/H ratio reflects the total number of C-D bonds relative to the total number of C-H bonds. To derive the D/H ratio of benzonitrile it is necessary to take into account the statistics of the different deuterated isotopologs, where the \textit{ortho} and \textit{meta} forms have two equivalent positions for the deuterium atom while the \textit{para} isomer has only one possible position for D. We assumed that the \textit{ortho}, \textit{meta}, and \textit{para} isomers have statistical abundance ratios, 2:2:1, respectively. In this case, the upper limits on the column densities of the \textit{ortho}, \textit{meta}, and \textit{para} isomers would be 3.0\,$\times$\,10$^{10}$ cm$^{-2}$, 3.0\,$\times$\,10$^{10}$ cm$^{-2}$, and 1.5\,$\times$\,10$^{10}$ cm$^{-2}$, respectively. Adopting the column density of the parent species from \cite{Cernicharo2021b}, we end up with a D/H ratio of $<$\,1.2\,\% for benzonitrile in \mbox{TMC-1}. The upper limit is about three orders of magnitude higher than the elemental D/H ratio in the solar neighborhood, in the range from 1.5\,$\times$\,10$^{-5}$ to 2.3\,$\times$\,10$^{-5}$ \citep{Linsky2006}.

Several singly deuterated molecules have been detected in \mbox{TMC-1} (see \citealt{Cabezas2021a,Cabezas2021b,Cabezas2022,Tercero2024,Cernicharo2024b}). These comprise the aliphatic hydrocarbons $c$-C$_3$H$_2$, CH$_3$CCH, C$_4$H, H$_2$C$_4$, and CH$_3$C$_4$H and the N-bearing carbon chain molecules CH$_2$CN, CH$_3$CN, HC$_3$N, HNC$_3$, HCCNC, CH$_3$C$_3$N, and HC$_5$N. The D/H ratios derived for these molecules are in the range of 0.06-3.3\,\% (see Table 4 of \citealt{Cabezas2022}, Table A.5 of \citealt{Tercero2024}, and Table 2 of \citealt{Cernicharo2024b}). The presence of these deuterated species is explained in terms of isotopic fractionation reactions that favor the incorporation of deuterium into molecules and that become efficient at the very low temperatures of cold dark clouds \citep{Roberts2004,Roueff2005}. The D/H ratio derived for benzonitrile, $<$\,1.2\,\%, is fully consistent with the range of D/H ratios derived for other monodeuterated molecules in \mbox{TMC-1}.

We were also interested in comparing the D/H ratio derived for benzonitrle in \mbox{TMC-1} with the ones derived for large unspecific PAHs in UV-illuminated interstellar media. In these regions, the vibrational modes of PAHs are excited by the absorption of UV photons. This results in the emission of several strong IR bands between 3.3\,$\mu$m and 19\,$\mu$m corresponding to the different C$-$H and C$-$C bending and stretching modes \citep{Draine2007}. The D/H ratio in these PAHs was measured using the intensity ratio between the 4.65\,$\mu$m band, associated with the C$-$D stretch in aliphatic sites, and the 3.4\,$\mu$m one, corresponding to the C$-$H stretch counterpart, together with the intensity ratio between the 4.35\,$\mu$m and 3.3\,$\mu$m bands, which correspond to the C$-$D and C$-$H stretches, respectively, at aromatic sites. Using observations with the Infrared Space Observatory (ISO), \cite{Peeters2004} reported high D/H ratios (distributed over the aromatic and aliphatic carbon atoms) of 17\,\%\,$\pm$\,3\,\% and 36\,\%\,$\pm$\,8\,\% in the PDRs Orion Bar and M\,17, respectively. However, using data from \textit{AKARI}, \cite{Onaka2014} derived smaller D/H ratios of 2-3\,\% in these two PDRs, and \cite{Doney2016} found evidence of deuterated PAHs in the spectra of only 6 of the 53 H{\tiny II} regions investigated, with variable D/H ratios in the range of 3-44\,\%.

More accurate D/H ratios of PAHs have recently been obtained using James Webb Space Telescope (JWST) observations. In general, it has been found that the D\slash H ratio for aliphatic side groups (D$_{\rm ali}$\slash H$_{\rm ali}$) is significantly larger than the one for aromatic ones (D$_{\rm aro}$\slash H$_{\rm aro}$). The aliphatic bands are thought to be mostly produced by methyl ($-$CH$_3$) and hydrogenated ($-$H) groups and their deuterated equivalents (e.g., \citealt{Schutte1993,Bernstein1996,Buragohain2020,Pla2020,Yang2023}). In the Orion Bar, \citet{Peeters2024} found D$_{\rm ali}$\slash H$_{\rm ali}$ ratios between 4.1\,\% and 22.5\,\%, while in the PDR M\,17, \citet{Boersma2023} measured 31\,$\pm$\,12.7\,\%. In the spiral galaxy M\,51, \cite{Draine2025} derived D$_{\rm ali}$\slash H$_{\rm ali}$ of 17\,$\pm$\,2\,\%, while in the disk of the local starburst galaxy NGC\,3256-S we derived a D$_{\rm ali}$\slash H$_{\rm ali}$ ratio of $\sim$12\,\% following the method described in \citet{Yang2023} and using the band intensity ratio $I_{4.65}\slash I_{3.4}$\,=\,0.10\,$\pm$\,0.01 observed in the JWST spectrum presented in \citet{Pereira-Santaella2024}. In contrast, the deuterated aromatic band at 4.35\,$\mu$m is undetected in M\,17 (D$_{\rm aro}$\slash H$_{\rm aro}$\,$<$\,0.5\,\%; \citealt{Boersma2023}), in NGC\,3256-S (D$_{\rm aro}$\slash H$_{\rm aro}$\,$<$\,4\,\% from the observed 3$\sigma$ upper limit, $I_{4.4}\slash I_{3.4}$\,$<$\,0.03), and in the M\,51 galaxy (D$_{\rm aro}$\slash H$_{\rm aro}$\,$<$\,1.6\,\%; \citealt{Draine2025}). For the Orion bar, \citet{Yang2025} report an upper limit of D$_{\rm aro}$\slash H$_{\rm aro}<$14\,\%, since the 4.35\,$\mu$m band might be contaminated by C$-$N stretches.

The high D$_{\rm ali}$\slash H$_{\rm ali}$ ratios given above can be misleading because in astronomical PAHs, H atoms at aliphatic C sites are less abundant than at aromatic C sites. Therefore, even if the D$_{\rm ali}$\slash H$_{\rm ali}$ ratios are relatively high, the total D\slash H ratio (including both aliphatic and aromatic) is substantially lower: between 3\,\% and $<$\,17\,\% in the Orion Bar, 2-3\,\% in M\,17 \citep{Yang2025}, 3-5\,\% in the spiral galaxy M\,51, and between 1\,\% and $<$\,5\,\% in the starburst galaxy NGC\,3256-S. The D/H ratio of $<$\,1.2\,\% determined here for C$_6$H$_5$CN in \mbox{TMC-1} lies just at the low edge of the range of total D/H ratios (including aliphatic and aromatic) derived in the four aforementioned UV-irradiated regions (between 1\,\% and $<$\,17\,\%). It probably lies below that range taking into account that it is a 3$\sigma$ upper limit, which would make us conclude that the level of deuteration of PAHs in \mbox{TMC-1}, probed by benzonitrile, is different from that in UV-illuminated clouds. However, the D atom in deuterated benzonitrile is bonded to an aromatic C, and thus we should compare with the D$_{\rm aro}$\slash H$_{\rm aro}$ ratios derived in the Orion Bar, M\,17, M\,51, and NGC\,3256-S, which are all upper limits ($<$\,14\,\%, $<$\,0.5\,\%, $<$\,1.6\,\%, and $<$\,4\,\%, respectively), since aromatic deuterium is not detected in any of these sources. In this regard, the level of deuteration of benzonitrile in \mbox{TMC-1} would be consistent with the D$_{\rm aro}$\slash H$_{\rm aro}$ ratios derived in UV-illuminated regions.

There are, however, further aspects to take into account. The PAHs detected in dark clouds have a distinctive feature compared to the ones in UV-illuminated regions. In \mbox{TMC-1}, the H atoms of PAHs are rarely in aliphatic sites, while in UV-irradiated regions H atoms in aliphatic sites represent a sizable fraction of the total number of H atoms, around 20\,\% \citep{Chiar2013,Yang2017,Hensley2020,Draine2025}. We can estimate the fraction of aliphatic H atoms in the PAHs detected to date in \mbox{TMC-1}. Since for many of them only the CN derivative is detected, we estimate the abundance of the pure hydrocarbon scaling up by a factor of 27, which is the abundance ratio determined for cyclopentadiene \citep{Cernicharo2021b,Cernicharo2022}. The estimated column densities, in units of 10$^{13}$ cm$^{-2}$, are thus 1.2 for cyclopentadiene (C$_5$H$_6$; \citealt{Cernicharo2021a}), 3.2 for benzene (C$_6$H$_6$; \citealt{Cernicharo2021b}), 1.6 for indene (C$_9$H$_8$; \citealt{Cernicharo2021a}), 3.9 for naphthalene (C$_{10}$H$_8$; \citealt{McGuire2021}), 5.1 for acenaphthylene (C$_{12}$H$_8$; \citealt{Cernicharo2024a}), 9.8 for pyrene (C$_{16}$H$_{10}$; \citealt{Wenzel2024}), and 7.3 for coronene (C$_{24}$H$_{12}$; \citealt{Wenzel2025}). The column density of C-H bonds is thus 3.0\,$\times$\,10$^{15}$ cm$^{-2}$. In the above molecules, most H atoms are at aromatic sites; only two H atoms in cyclopentadiene and two in indene are at aliphatic C sites of the hydrogenated type. The column density of H atoms in hydrogenated aliphatic C sites is therefore 5.6\,$\times$\,10$^{13}$ cm$^{-2}$, which implies a fraction of hydrogenated aliphatic H atoms of 1.9\,\%. Methyl groups are also thought to be an important contribution to the aliphatic fraction of H atoms in large PAHs. However, no methylated PAH has been detected in \mbox{TMC-1}. The simplest such molecule would be toluene (C$_6$H$_5$CH$_3$), for which we estimate a 3$\sigma$ upper limit on its column density of 4\,$\times$\,10$^{12}$ cm$^{-2}$. The column density of H atoms at methyl aliphatic C sites would be $<$\,1.2\,$\times$\,10$^{13}$ cm$^{-2}$, which implies a fraction of methyl aliphatic H atoms of $<$\,0.4\,\%. Therefore, the fraction of aliphatic H atoms (hydrogenated plus methyl) in \mbox{TMC-1} is estimated to be $<$\,2.3\,\%, which is substantially smaller than the values inferred for large PAHs in UV-irradiated regions. Moreover, given the low fraction of aliphatic H in \mbox{TMC-1}, aliphatic D probably contributes little compared to aromatic D, and we can thus view the aromatic D/H ratio of $<$\,1.2\,\% derived from benzonitrile as a proxy of the total D/H ratio.

In summary, the low deuteration level seen for benzonitrile in \mbox{TMC-1} is consistent with the range of values found for other monodeuterated molecules in this same cloud, although it is on the low edge of (and very likely below) the range of D/H values derived for large PAHs in the PDRs Orion Bar and M\,17 and in the galaxies M\,51 and NGC\,3256-S. That is, the population of PAHs in dark clouds does not seem to share the deuterium enrichment seen in PDRs. If we assume that such deuterium enhancement is also present in the large PAHs observed in diffuse clouds, the conclusion would be that PAHs in dark clouds are not inherited from the previous diffuse cloud stage. However, it is not fully clear whether this deuterium enrichment holds for PAHs in all kinds of interstellar regions. \cite{Peeters2024} argue that the D enhancement may occur in situ at the border of PDRs due to UV radiation and/or density, although it is still unknown what mechanism is behind deuterium enrichment of PAHs and whether it is a local effect that occurs in certain PDR-like regions or whether it is a characteristic of PAHs in different kinds of UV-irradiated regions, diffuse clouds included.

\section{Conclusions} \label{sec:conclusions}

We have characterized in the laboratory the rotational spectra of the three isomers (\textit{ortho}, \textit{meta}, and \textit{para}) of monodeuterated benzonitrile in the 2-18 GHz and 75-110 GHz frequency ranges. The measured frequencies have been used to predict accurate transition frequencies in the Q band and search for these three isomers in the cold dark cloud \mbox{TMC-1} using data from the QUIJOTE line survey. We did not detect any of the three species, and we have derived a 3$\sigma$ upper limit on the column density of each of the three deuterated isomers of benzonitrile of 3.0\,$\times$\,10$^{10}$ cm$^{-2}$, which implies a D/H ratio of $<$\,1.2\,\%. This value is in line with the range of D/H ratios observed for smaller molecules in \mbox{TMC-1} from radioastronomical data (0.06-3.3\,\%) and below the range of D/H values inferred for large PAHs from infrared observations of UV-illuminated regions (between 1\,\% and $<$\,17\,\%). Although drawing a clear link between deuteration of PAHs in dark clouds and UV-irradiated regions is not straightforward, our results suggest that PAHs in cold dark clouds are not formed from the fragmentation of large PAHs inherited from a previous diffuse phase in a top-down scenario.\\

\noindent \textbf{Data availability}\\

\noindent The measured and analyzed rotational transitions and the frequency predictions for each of the three isomers of monodeuterated benzonitrile are available at \href{https://zenodo.org/records/15403336}{zenodo}.

\begin{acknowledgements}

We acknowledge funding support from Spanish Ministerio de Ciencia, Innovaci\'on, y Universidades through grants PID2021-125015NB-I00 (C.P. and A.L.), PID2022-139933NB-I00 (D.P. and J.J.), PID2023-146667NB-I00 (M.P.S. and J.R.G), and PID2023-147545NB-I00 (M.A., C.C., and J.C.). A.L.S. and M.P.S. acknowledge funding support from grants RYC2022-037922-I (A.L.S.) and RYC2021-033094-I and CNS2023-145506 (M.P.S.), funded by MCIN/AEI/10.13039/501100011033, the FSE+, and the European Union NextGenerationEU/PRTR. C.P. thanks the European Research Council for the CoG HydroChiral (Grant Agreement No 101124939). A.L.S., C.P., and A.L. also thank funding from Junta de Castilla y Le\'on, Grant INFRARED IR3032-UVA13. D.P. and J.J. thank Xunta de Galicia and the European Regional Development Fund (ERDF) for grant ED431G 2023/03 (Centro de Investigaci\'on do Sistema Universitario de Galicia accreditation 2023-2027). I.G.B. is supported by the Programa Atracci\'on de Talento Investigador \textit{C\'esar Nombela} via grant 2023-T1/TEC-29030 funded by the Community of Madrid. This work is based in part on observations made with the NASA/ESA/CSA James Webb Space Telescope. The data were obtained from the Mikulski Archive for Space Telescopes at the Space Telescope Science Institute, which is operated by the Association of Universities for Research in Astronomy, Inc., under NASA contract NAS 5-03127 for JWST; and from the European JWST archive (eJWST) operated by the ESAC Science Data Centre (ESDC) of the European Space Agency. These observations are associated with program \#1328. We thank the anonymous referee for a constructive report that helped to improve this manuscript.

\end{acknowledgements}

\begin{appendix}

\onecolumn

\section{Synthesis of \textit{ortho}, \textit{meta}, and \textit{para} deuterated benzonitrile} \label{app:synthesis}

\subsection{General methods}

All reactions were carried out under argon using oven-dried glassware. Anhydrous dimethyl sulfoxide (DMSO) was taken from a MBraun SPS-800 Solvent Purification System. 2-bromobenzonitrile, 3-bromobenzonitrile 4-bromobenzonitrile were purchased from BLD Pharmatech, and sodium formate from Biosynth, and were used without further purification. Thin-layer chromatography (TLC) was performed on Merck silica gel 60 F254 and chromatograms were visualized with UV light (254 and 360 nm). Column chromatography was performed on Merck silica gel 60 (ASTM 230-400 mesh). $^1$H and $^{13}$C nuclear magnetic resonance (NMR) spectra were recorded at 300 and 75 MHz (Varian Mercury-300 instrument). GC-MS experiments were conducted using a HP 5973 INERT series and an Agilent HP-5MS.

\subsection{Experimental procedures and characterization data}

\begin{figure}[h!]
\centering
\includegraphics[angle=0,width=0.4\textwidth]{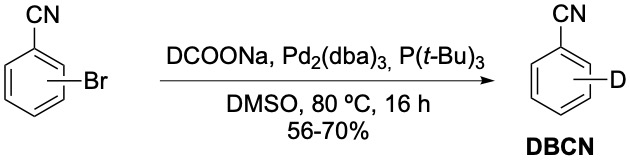}
\caption{General procedure for the synthesis of \textit{ortho}, \textit{meta}, and \textit{para} deuterated benzonitrile.} \label{fig:synthesis}
\end{figure}

The synthesis of the three isomers of deuterated benzonitrile (DBCN in Fig.\,\ref{fig:synthesis}) was done according to an adapted procedure from the literature (\citealt{Zhang2015}; see Fig.\,\ref{fig:synthesis}).

We added Pd$_2$(dba)$_3$ (50 mg, 0.055 mmol), DCOONa (380 mg, 5.5 mmol) in dry DMSO (3 mL), and P($t$-Bu)$_3$ (33.3 mg, 0.165 mmol) to a stirred solution containing the corresponding bromobenzonitrile, 2-, 3-, or 4- (500 mg, 2.74 mmol). The mixture was heated at 80$^{\circ}$C until GC-MS showed the full conversion of the starting material. After that, the reaction was quenched with saturated NH$_4$Cl, extracted with CH$_2$Cl$_2$, and the organic phase was dried over Na$_2$SO$_4$, filtered and concentrated under high vacuum. The crude product was purified by column chromatography (SiO$_2$, hexane) yielding the corresponding deuterated benzonitrile, \textit{ortho}, \textit{meta}, or \textit{para}. The $^1$H and $^{13}$C NMR spectra are available at \href{https://zenodo.org/records/15403336}{zenodo}.

\textbf{Benzonitrile-2-d (\textit{ortho} deuterated benzonitrile)}: yield: 70\%. Clear oil. \textbf{$^1$H-RMN} (300 MHz, CDCl$_3$) $\delta$: 7.68 – 7.56 (m, 2H), 7.46 (ddd, $J$ = 7.9, 5.5, 2.6 Hz, 2H) ppm. \textbf{$^{13}$C-RMN-DEPT} (75 MHz, CDCl$_3$) $\delta$: 132.77 (CH), 132.21 (CH), 132.14 (t, $J$ = 25.62 Hz, CD), 129.13 (CH), 129.02 (CH), 118.81 (CN), 112.40 (C) ppm.

\textbf{Benzonitrile-3-d (\textit{meta} deuterated benzonitrile)}: yield: 70\%. Clear oil. \textbf{$^1$H-RMN} (300 MHz, CDCl$_3$) $\delta$: 7.68 – 7.58 (m, 3H), 7.50 – 7.43 (m, 1H) ppm.\textbf{$^{13}$C-RMN-DEPT} (75 MHz, CDCl$_3$) $\delta$: 132.83 (CH), 132.27 (t, $J$ = 23.6 Hz, CD), 132.16 (CH), 132.06 (CH), 129.18 (CH), 118.82 (CN), 112.51 (C) ppm.

\textbf{Benzonitrile-4-d (\textit{para} deuterated benzonitrile)}: yield: 56\%. Clear oil. \textbf{$^1$H-RMN} (300 MHz, CDCl$_3$) $\delta$: 7.68 – 7.63 (d, $J$ = 7.6 Hz, 2H), 7.47 (d, $J$ = 7.7 Hz, 2H) ppm. \textbf{$^{13}$C-RMN-DEPT} (75 MHz, CDCl$_3$) $\delta$: 132.48 (t, $J$ = 22.7 Hz CD), 132.2 (2xCH), 129.0 (2xCH), 118.9 (CN), 112.4 (C) ppm.

\end{appendix}

\end{document}